\documentclass[12pt]{article}
\usepackage[dvips]{graphics}
\usepackage{amssymb}
\usepackage{amsmath}

\begin{document}



\begin{center}
{\large \bf Bound on the tau neutrino magnetic moment\\
 from the Super-Kamiokande data} 
\end{center}
\vspace{0.5cm}
\begin{center}
{\large  
S.N.Gninenko\footnote{ Sergei.Gninenko\char 64 cern.ch}
}
\end{center}

\begin{center}
{\it CERN, CH-1211 Geneva 23, Switzerland\\
and\\
Institute for Nuclear Research, RAS , Moscow 117312, Russia}  
\end{center}

\begin{abstract}
It is shown that recent results from the Super-Kamiokande detector 
 constrain the tau neutrino diagonal magnetic moment to 
$\mu_{\nu_{\tau}} < 1.3\times 10^{-7} \mu_{B}$ for the case of  
$\nu_{\mu}\rightarrow \nu_{\tau}$ interpretation of the atmospheric 
neutrino anomaly.\ It is pointed out that the large magnetic moment of the 
tau neutrino could affect further 
understanding of the origin of the anomaly.
\end{abstract}

\vspace{0.5cm}

Recent results from the Super-Kamiokande (S-K) detector  
give evidence for neutrino oscillations \cite{1} - \cite{4}.\ 
One of the favourable interpretations of the atmospheric neutrino anomaly 
suggested by the S-K collaboration is related to the existence of  
$\nu_{\mu}\rightarrow \nu_{\tau}$ neutrino oscillations.\
This 
implies that tau neutrinos 
have  non-zero masses and therefore  may also have 
a non-zero diagonal magnetic moment ( see e.g. ref. \cite{5}).\ 
As a consequence,   
 massive tau neutrinos could manifest themselves in terrestrial 
experiments through the effect of 
$\nu_{\mu}\rightarrow \nu_{\tau}$
neutrino oscillations, and, if the magnetic moment value is large enough,
through tau neutrino  electromagnetic interactions.

 In ref.\cite{6} (see also \cite{7}), it was shown that the combined existence 
of  $\nu_{\mu}\rightarrow \nu_{\tau}$ (and/or $\nu_{e}\rightarrow \nu_{\tau}$)
 oscillations and  non-zero magnetic moment of the tau neutrino would 
increase 
the total rate of events in $\nu_{\mu}(\nu_{e})$ neutrino- electron scattering experiments.\ It was used to constrain the mixing angles of the tau 
neutrino with neutrinos of another flavour.\ In this Letter it is shown 
that this effect can also be used to constrain the magnetic moment 
of the tau neutrino from the S-K atmospheric neutrino data.\

Assuming 
 that a muon neutrino beam has a component of tau neutrinos due to 
$\nu_{\mu}\rightarrow \nu_{\tau}$ oscillations, 
in the case of two-neutrino mixing,
 neutrino states  evolve with time $t$ as  

\begin{equation}
|\nu(t)> = a(t)|\nu_{\mu}>  + b(t)|\nu_{\tau}, \mu_{\nu_\tau}\neq 0>
\end{equation}    

where $|\nu_{\mu}>$ and $|\nu_{\tau}>$ denote  weak eigenstates 
of $\nu_{\mu}$ and $\nu_{\tau}$ neutrinos, and $a^{2}(t),~b^{2}(t)$ are the 
probabilities of finding $\nu_{\mu}$ or $\nu_{\tau}$ in the beam at a given moment
 $t$, respectively.\ It is assumed that $a^{2}(0) =1$ at $t=0$.\  
The probability $b^{2}(t)$ 
depends on the parameters of $\nu_{\mu} - \nu_{\tau}$ oscillations as \cite{8}:   

\begin{equation}
b^{2}(t\simeq \frac{L}{c}) = P(\nu_{\mu}\rightarrow\nu_{\tau}) = 
sin^{2}2\theta_{\mu\tau} sin^{2}\frac{\Delta m^{2} L}{4E_{\nu}} 
\end{equation}
or
\begin{equation}
P(\nu_{\mu}\rightarrow \nu_{\tau}) \approx sin^{2}2\theta_{\mu\tau} sin^{2}\frac{1.27 \Delta m^{2}(eV^{2}) L(km)}{E_{\nu}(GeV)}
\end{equation}

where $sin^{2}2\theta_{\mu\tau}$ is the
 mixing angle,
$\Delta m^{2} = \bigl|m^{2}_{3} - m^{2}_{2}\bigr|$ 
 is the difference of the squares of the mass eigenstates in $eV^{2}$, $E_{\nu}$ is the neutrino energy in GeV, and $L$ is
the mean distance between the neutrino source and the detector in km.\ 
In the above formula it is also assumed that 
 magnetic field $B$ is weak enough not to affect the probability of
oscillations $P(\nu_{\mu}\rightarrow \nu_{\tau})$, i.e. \cite{9}:

\begin{equation}
 \Delta m^{2}/2E_{\nu} \gg \mu_{\nu_\tau}B
\end{equation}

Then if the magnetic moment of the
 $\nu_{\tau}$ exists,
it will contribute to a non-coherent part
of the $\nu_{\tau}e^{-}$ scattering cross section 
via the reaction that changes the helicity of the tau neutrino (hence 
 right-handed neutrino states should exist).\  
 This 
might result in a contribution to the observed deviations from unity of
the flavour ratio  of the atmospheric neutrino flux, 
$R \equiv \bigl((N_{\nu_{\mu}}+N_{\overline{\nu}_{\mu}})/
(N_{\nu_e}+N_{\overline{\nu}_{e}})\bigr)_{data}/\bigl((N_{\nu_{\mu}}+N_{\overline{\nu}_{\mu}})/
(N_{\nu_e}+N_{\overline{\nu}_{e}})\bigr)_{MC}$, in the S-K detector for both 
data and Monte Carlo (MC) simulation.\ 
Here,  $N_{\nu_{\mu},\overline{\nu}_{\mu}}$ and $N_{\nu_e,\overline{\nu}_e}$
are the number of muon -like ($\mu$-like) and electron-like ($e$-like)
fully-contained events in the S-K detector \cite{1}.\ 

 Indeed, since the electromagnetic cross section is orders of 
magnitude larger than the weak cross section, even a small fraction of 
tau neutrinos with non-zero magnetic moment in the atmospheric  neutrino flux
 could lead
 to an observable excess of isolated electrons in atmospheric neutrino interactions in the S-K
detector whose signature is identical to that of e-like events.
 Note that the magnetic moments 
of muon or electron  neutrino are experimentally proved to be too small  
to give a noticeable contribution to the neutrino interaction rate.\ 
 The production rate of isolated electrons via 
$\nu_{\tau} e^{-}$  scattering in 
the detector depends  on the probability 
$P(\nu_{\mu}\rightarrow\nu_{\tau})$ of finding  $\nu_{\tau}$ neutrinos in the 
atmospheric neutrino flux.\ This probability can be calculated from the neutrino 
survival and transition probabilities using Eq.(2).\

The neutrino-electron scattering process via magnetic moment has a cross 
section \cite {10}:
 
\begin{equation}
\frac{d \sigma_{\mu}}{dE_e} = \frac{\pi \alpha^{2}}{m_{e}^{2}} 
\frac{\mu_{\nu}^{2}}{\mu_{B}^{2}}\Bigl(\frac{1}{E_e} - \frac{1}{E_{\nu}}\Bigr) 
\end{equation}
 
where $E_{e}$ is the  electron energy, 
$\mu_{\nu}$ is the neutrino magnetic moment and $\mu_B = e/2 m_e$ is Bohr magneton.\ For the case of constant $E_{\nu}$ the integral cross section is 

\begin{equation}
\begin{split}
&\sigma_{\mu} = \frac{\pi \alpha^{2}}{m_{e}^{2}} 
\frac{\mu_{\nu}^{2}}{\mu_{B}^{2}}\Bigl[ln\Bigl(\frac{E_e^{max}}{E_e^{min}}\Bigr) - \frac{(E_e^{max} - E_e^{min})}{E_{\nu}}\Bigr]\\
& = 2.7 \times 10^{-39}\Bigl(\frac{\mu_{\nu}}{10^{-7}\mu_{B}}\Bigr)^2 
\Bigl[ln\Bigl(\frac{E_e^{max}}{E_e^{min}}\Bigr) - \frac{(E_e^{max} - E_e^{min})}{E_{\nu}}\Bigr]~ cm^2 
\end{split}
\end{equation}

where $E_e^{max},~ E_e^{min}$ are the high and low electron energy cuts, 
respectively.\ 
Note that the integral cross-
section $\sigma_{\mu}$ depends very weakly  on the neutrino energy.\ It 
rises only logarithmically with the neutrino energy, while the total neutrino 
cross section rises linearly with $E_{\nu}$.\ Thus, it is advantageous to 
search for a tau neutrino magnetic moment using the low energy
(sub-GeV, $E_{visible} < 1.33$ GeV) S-K data \cite{1}.\ 
 
Approximately, all atmospheric neutrino giving rise to the fully 
contained $\mu$-like events in the S-K detector have energy above  400 MeV, see e.g. 
Ref. \cite{11}
(MC studies showed the mean neutrino energy for CC interactions
to be about 800 MeV for $\mu$-like events \cite{1}).\ 
Thus, for the lowest  electron energies analysed in the S-K, from  $E_e^{min}=$ 100 MeV to 
$E_e^{max}=$ 200 MeV (first bin of the histogram in Fig.4(a) 
corresponding to e-like events, ref. \cite{1}),
the cross section of $\nu_{\tau} - e$ scattering due to non-zero magnetic moment is 

\begin{equation}
\sigma_{\mu} = 1.2 \times 10^{-39} (\mu_{\nu}/10^{-7}
\mu_B)^2~cm^2
\end{equation}

Here, we assume the tau neutrino energy $E_{\nu}=400$ MeV, thus the integral cross section is underestimated.
   
The total number $\Delta N_e$  of e-like events from 
$\nu_{\tau}-e$ scattering in the first energy of Fig.4(a) in \cite{1} can be written in the form 
\begin{equation}
\Delta N_e = \int k_0\cdot f_{\nu}\cdot \sigma_{\mu}\cdot \varepsilon \cdot d\Omega\cdot dE_{\nu}
=  k_1 N_{\nu_{\tau}}  \sigma_{\mu}
\end{equation}
 
where $k_0$ is a factor related to the $\nu$-target mass, $f_{\nu}$ is tau neutrino flux, the cross section $\sigma_{\mu}$ is constant, 
 $\varepsilon$ is detection efficiency which is  
 practically independent of energy for $E_e > 100$ MeV \cite{1},
$k_1$ is a factor corresponding to the convolution of detector acceptance, detection efficiency and $\nu$-target mass and $N_{\nu_{\tau}}$ is the total
number of tau-neutrinos crossing fiducial volume of the S-K detector.\ 

The number of tau-neutrinos is estimated using 
 the number of the {\it initially} produced muon atmospheric neutrino:

\begin{equation}
 N_{\nu_{\tau}} = (N_{\nu_{\mu}}+N_{\overline{\nu}_{\mu}})
\overline{P}(\nu_{\mu} \rightarrow \nu_{\tau}) 
\end{equation}

The observation of the small value 
of $R = 0.61 \pm 0.03(stat.) \pm 0.05(syst.)$ \cite{1} and zenith angle dependent deficit of $\mu$-like events suggests, that 
in the case of two -neutrino oscillation scenario of $\nu_{\mu}\rightarrow \nu_{\tau}$ more than 30$\%~ (90\% C.L.)$ of the initially produced muon neutrinos arrive at
the S-K detector as tau neutrinos \cite{2,3}. We assume that the average probability of 
oscillations $\overline{P}(\nu_{\mu} \rightarrow \nu_{\tau}) = 0.3$.\
  
Muon-neutrino quasi-elastic events detected in the S-K detector were used to estimate the expected 
total number of muon neutrino.\ Similarly to Eq.(8) one can write 

\begin{equation}
\begin{split}
& N_{\mu}^{q.e.} = k_2 (N_{\nu_{\mu}}\sigma_{q.e.}^{\nu}  + 
N_{\overline{\nu}_{\mu}}\sigma_{q.e.}^{\overline{\nu}}) 
\Bigl(1 - \overline{P}(\nu_{\mu} \rightarrow \nu_{\tau})\Bigr)\\
& = k_2 (N_{\nu_{\mu}} + N_{\overline{\nu}_{\mu}})
\Bigl(1 - \overline{P}(\nu_{\mu} \rightarrow \nu_{\tau})\Bigr)\overline{\sigma}_{q.e.}
\end{split}
\end{equation}

where $k_2$ has the same meaning as $k_1$ in Eq.(8), and 
$\sigma_{q.e.}^{\nu}$,$\sigma_{q.e.}^{\overline{\nu}}$ are the 
cross-sections of 
quasi-elastic scattering of muon neutrino on (bounded) neutrons and anti-muon
neutrino on protons in the 
H$_2$0-target, respectively, which  were taken to be 
constant and to be  equal to its maximal values of $\sigma_{q.e.}^{\nu} = 1.0\times 10^{-38}~cm^2$ and $\sigma_{q.e.}^{\overline{\nu}} = 0.4 \times 10^{-38}~cm^2$ in the 
energy region $E_{\nu} <$1.33 GeV, see e.g. \cite{11}.\ 
The average cross section $\overline{\sigma}_{q.e.}$ = ($\sigma_{q.e.}^{\nu} +
\sigma_{q.e.}^{\overline{\nu}})$/2, since according to  MC simulation used in the experiment 
the ratio  $N_{\nu_{\mu}}/N_{\overline{\nu}_{\mu}}$ = 1 within a few percent
for the neutrino energy range considered \cite{12}.\
Thus, the total muon neutrino 
flux is underestimated.\ 
The number of detected 
quasi-elastic muon-neutrino events can be extracted from the number of
${\mu}$-like events in Table 1 of ref.\cite{1}.\ Under the 
$\nu_{\mu} \rightarrow \nu_{\tau}$ oscillation hypothesis used the fractions 
of  $\nu_{e}CC,~ \nu_{\mu}CC$ and $NC$ events in a detected sample 
of $\mu$-like events calculated for $\overline{P}(\nu_{\mu} \rightarrow \nu_{\tau}) = 0.3$
should be $0.7\%,~94\%$ and $5.3\%$, respectively, 
instead of $0.5\%,~95.8\%$ and $3.7\%$ given in Table 1 of ref.\cite{1}.\ 
Thus, $N_{\mu}^{q.e.} =900\times .94\times 916.9/1166.5 = 665$ events.\ 
We also assume that the final state signature in water
Cherenkov detectors are practically the same for 
quasi-elastic $\nu_{e}$ and $\nu_{\tau} -e $ elastic scattering processes,
so detection efficiencies for these processes are the same.\
The efficiencies for identifying quasi-elastic $\nu_e$ and $\nu_{\mu}$ events
were 93$\%$ and 95$\%$, respectively \cite{1}.\ Finally, since detection efficiences for fully-contained e-like ($P_e >$100 MeV/c) and $\mu$-like ($P_{\mu} >$
200 MeV/c) events are practically energy independent we neglect this 
small difference in efficiencies and  assume that $k_1 = k_2 $.\

Using the above value for $N_{\mu}^{q.e.}$, Eqs.(7-10)
 and $\mu_{\nu_{\tau}}=5.4\times 10^{-7}\mu_B$,
 which corresponds to the BEBC experiment upper limit
on diagonal tau neutrino magnetic moment \cite{13}, it is found
that in the first bin of histogram in Fig.4(a) \cite{1} the S-K experiment 
should detect:

\begin{equation}
\Delta N_e^1 = N_{\mu}^{q.e.}\cdot  
\frac{\overline{P}(\nu_{\mu} \rightarrow \nu_{\tau})}{1 - 
\overline{P}(\nu_{\mu} \rightarrow \nu_{\tau})}\cdot 
\frac{\sigma_{\mu}}{\overline{\sigma}_{q.e.}} \simeq 1420~ events
\end{equation}

This number is much greater than the number of observed events, 
$N_{e,data}^1 = 160$ events, or the number of events predicted by MC simulation, 
$N_{e,MC}^1 = 125$ events in this energy bin.\ {\footnote { These numbers were read off Fig.4(a) of ref.\cite{1}}}

The bound for the tau neutrino magnetic moment can be  calculated by using the 
following relation :

\begin{equation}
N_{\mu}^{q.e.}\cdot 
\frac{\overline{P}(\nu_{\mu} \rightarrow \nu_{\tau})}{1 - 
\overline{P}(\nu_{\mu} \rightarrow \nu_{\tau})}\cdot 
\frac{\sigma_{\mu}}{\overline{\sigma}_{q.e.}} < (\Delta N_e^1)^{90} 
\end{equation}

where  $(\Delta N_e^1)^{90}$ = $(N_{e,data}^1 - N_{e,MC}^1)^{90}$
($\simeq$ 80 events) is the 90$\%~C.L.$ upper limit for the expected number of 
e-like events from $\nu_{\tau} - e $ scattering in the first energy bin of 
Fig.4(a) (ref.\cite{1}), calculated taken 
into account the systematic uncertainty in the absolute normalisation of the 
MC events of 25$\%$ \cite{1} by added the errors in quadrature to the 
statistical error.\  
 
This results in a conservative bound

\begin{equation}
\mu_{\nu_{\tau}} < 1.3 \times 10^{-7} \mu_B
\end{equation}

The bound is a factor 4 better than the previously published
bound obtained by the BEBC experiment \cite{13} and is obtained on the  
assumption that $\nu_{\mu} \rightarrow \nu_{\tau}$ oscillations are the origin 
of the anomaly in atmospheric neutrino data observed by the S-K 
experiment.\ It can be improved by a more detailed analysis of the S-K data.\
Note that Eq.(4) and limit of Eq.(13) are consistent for the region of
$\Delta m^2 \sim (10^{-3} -10^{-2})$ eV$^2$ suggested by the S-K analysis
\cite{4}, $E_{\nu} \simeq O$(1 GeV) and for B equal to the average magnetic
field of the earth (B$\lesssim$1 Gauss).  

   In the S-K detector, the oscillation scenario was checked
by using $\pi^0$ events \cite{1}.\ For $\nu_{\mu}\rightarrow\nu_{\tau}$ 
oscillations the number of neutral -current (NC) events should be unchanged 
by neutrino oscillations.\ For $\nu_{\mu}\rightarrow\nu_{\tau}$ oscillations
the number of $\nu_e CC$ events ( e-like events) should also be unchanged 
by neutrino oscillations.\ Therefore, the $(\pi^0/e)$ ratio of the data 
should agree with the same ratio of the Monte Carlo without oscillations for 
$\nu_{\mu}\rightarrow\nu_{\tau}$ case \cite{4}.\ It was obtained 
 $(\pi^0/e)_{data}$/$(\pi^0/e)_{MC}$ = $0.93\pm0.07(stat)\pm0.19(syst)$
( preliminary).\ The result is consistent with the   
 $\nu_{\mu} \rightarrow \nu_{\tau}$ interpretation of the S-K data, however 
it cannot exclude $\nu_{\mu}\rightarrow\nu_{s}$ 
oscillation hypothesis completely. 

In a recent publication \cite{14}, it was pointed out that a large diagonal
and/or transition magnetic moment of neutrino can contribute to the neutral 
current effects ( by a single $\pi^0$ production) used to distinguish the mechanisms of muon neutrino 
oscillation to tau neutrino or to a sterile neutrino.\ The effect discussed 
in the present Letter, as well as the effect discussed in ref. \cite{14},
definitely can affect the interpretation of atmospheric 
neutrino data.\    

Finally note that helicity flipped states of neutrinos with the large magnetic
 moment would be trapped in the SN1987a core, so there is no limit on the 
neutrino magnetic moment form SN1987a \cite{15}.\ Others astrophysical 
constraints on $\mu_{\nu_{\tau}}$ are also model dependent and have 
considerable theoretical and 
experimental uncertainties.    

\vspace{1.0cm}

{\large \bf Acknowledgements}\\

I thank N.V.~Krasnikov and V.A.~Matveev for useful discussions.

\end{document}